\documentclass[acmsmall,nonacm]{acmart}
\usepackage[utf8]{inputenc}
\usepackage[T1]{fontenc}
\usepackage{amsmath}
\usepackage{textcomp}
\usepackage{capt-of}
\usepackage{hyperref}
\usepackage{newunicodechar}
\usepackage{pmboxdraw}
\RequirePackage{fancyvrb}
\usepackage{anyfontsize}
\usepackage{pifont}
\usepackage{fancyvrb}
\usepackage[shortcuts]{extdash}

\newenvironment{Snippet}{\Verbatim[samepage=true,fontsize=\small]}{\endVerbatim}

\DefineVerbatimEnvironment{verbatim}{Verbatim}{samepage=true}
\newunicodechar{─}{\textSFx}
\newunicodechar{│}{\textSFxi}
\newunicodechar{╭}{\textSFi}
\newunicodechar{╰}{\textSFii}
\newunicodechar{╮}{\textSFiii}
\newunicodechar{╯}{\textSFiv}
\newunicodechar{═}{\textSFxliii}
\newunicodechar{║}{\textSFxxiv}
\newunicodechar{╔}{\textSFxxxix}
\newunicodechar{╚}{\textSFxxxviii}
\newunicodechar{╗}{\textSFxxv}
\newunicodechar{╝}{\textSFxxvi}
\newunicodechar{◀}{$\triangleleft$}
\newunicodechar{▶}{$\triangleright$}
\newunicodechar{▮}{\hspace{0.6pt}|\hspace{0.6pt}}
\author{Panicz Maciej Godek}
\email{godek.maciek@gmail.com}

\date{\today}

\setcopyright{acmcopyright}
\copyrightyear{2024}
\acmYear{2024}

\acmConference[Scheme'24]{the 24th Scheme Workshop}{September 7 2024}{Milan, Italy}

\setcopyright{rightsretained}

\title{An Implementation of a Visual Stepper in the GRASP Programming System}

\hypersetup{
 pdfauthor={Panicz Maciej Godek},
 pdftitle={An Implementation of a Visual Stepper in the GRASP programming system},
 pdfkeywords={},
 pdfsubject={},
 pdfcreator={Emacs 27.1 (Org mode 9.3)}, 
 pdflang={English}}
\begin{document}
\pagenumbering{gobble}

\begin{abstract}
The direct purpose of this paper -- as its title suggests -- is to
present how the visual evaluator extension is implemented in the
GRASP\footnote{GRASP is an open source project, available at
\url{https://github.com/panicz/grasp}} programming system.

The indirect purpose is to provide a tutorial around the design
of GRASP, and in particular -- around the architecture of its extension
mechanism.

Neither GRASP nor its extension mechanisms are, at the moment of
writing this paper, final or complete, and we are certain that some
details of the solutions described in here will change even before the
first release.

What will not change, though, is the set of problems that need to be
solved in order to build a system with capabilities similar to those
of GRASP. We believe that these problems might be of interest to the
Scheme community.

\end{abstract}

\ccsdesc[500]{Software and its engineering~Integrated and visual development environments}
\ccsdesc[300]{Human-centered computing~Visualization toolkits}
\ccsdesc[300]{Human-centered computing~Ubiquitous and mobile computing systems and tools}
\ccsdesc[100]{Computing methodologies~Graphics input devices}

\keywords{visual stepper, interactive programming,
structural editing}

\maketitle

\section{Introduction}

GRASP \cite{Godek2023} is a nascent graphical development environment
for the Scheme programming language, consisting of a structural editor
for S-expressions, as well as an extension mechanism, which allows to
display various forms of data in an arbitrary visual way.

This paper presents an implementation of one such extension, namely --
a visual stepper, which allows to observe single step reductions of a
purely functional subset of Scheme.

GRASP is currently available as a desktop application, a terminal
application and as an Android application.  It is implemented in Kawa,
which is a dialect of Scheme that runs on the JVM and provides
extensions for interfacing with JVM classes and libraries, including
the capability of defining new classes, as well as optional checked
monomorphic type
annotations\footnote{\url{https://www.gnu.org/software/kawa/}}.

\section{The fundamental ideas of GRASP}

At the core of GRASP, there are two fundamental ideas.  The first one
is that it is a generic editor for S-expressions that uses boxes to
represent a pair of matching parentheses.

This is what the definition of the factorial function looks like
in the GRASP representation:

\begin{Snippet}
╭        ╭     ╮                      ╮
│ define │ ! n │                      │
│        ╰     ╯                      │
│   ╭    ╭        ╮                 ╮ │
│   │ if │ <= n 1 │                 │ │
│   │    ╰        ╯                 │ │
│   │                               │ │
│   │       1                       │ │
│   │                               │ │
│   │       ╭     ╭   ╭       ╮ ╮ ╮ │ │
│   │       │ * n │ ! │ - n 1 │ │ │ │ │
╰   ╰       ╰     ╰   ╰       ╯ ╯ ╯ ╯ ╯
\end{Snippet}

The boxes can be manipulated using pointing interfaces such as touch
screens and computer mice. The left edge of a regular box can be used for
moving it to another place in the document (or some other document),
removing it, or copying (by using two presses). The right edge
of the box can be used for resizing it or splicing its contents
into its parent box.

The second idea is that the user is allowed to define custom boxes
that can be rendered and interacted with in special ways: a box then
receives an area in the document to which it can draw; it also
receives touch events from that area, and if the custom box is in the
focus, it additionally receives events from the keyboard.

Currently, GRASP contains a few predefined extensions. Among them,
there is the extension called \texttt{Button}, which reacts to presses by
invoking the thunk that was provided during its creation.

A button is created in a GRASP document by inputting an expression
such as

\begin{Snippet}
(Button label: "Press me" 
	action: (lambda () (WARN "Button pressed")))
\end{Snippet}

and then either "enchanting" the expression (by pressing the
\textit{tab} key with the text cursor positioned on either opening
or closing parenthesis of the outermost expression),
or evaluating it (by pressing \textit{ctrl+e}).

The result will be displayed roughly like this:

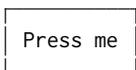
\begin{Snippet}
╭──────────╮
│ Press me │
╰──────────╯
\end{Snippet}

When the button is pressed, it causes the text "Button pressed" to be
displayed somewhere in the application log.

\section{The Visual Stepper Extension}

Another built-in extension is the visual stepper, which is the subject
of this work. It can be instantiated by typing \texttt{(Stepper
  \textit{<expression>})} into the editor, and pressing the tab key on
the closing parenthesis of the expression. For example, if
\texttt{\textit{<expression>}} is \texttt{(! 5)}, then the
corresponding stepper will look something like this:

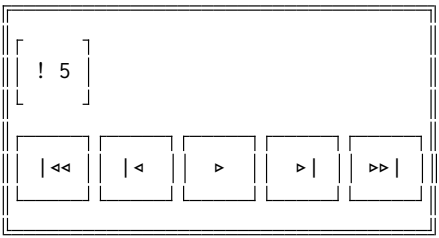
\begin{Snippet}
╔═══════════════════════════════════╗
║╭     ╮                            ║
║│ ! 5 │                            ║
║╰     ╯                            ║
║╭─────╮╭─────╮╭─────╮╭─────╮╭─────╮║
║│ ▮◀◀ ││ ▮◀  ││  ▶  ││  ▶▮ ││ ▶▶▮ │║
║╰─────╯╰─────╯╰─────╯╰─────╯╰─────╯║
╚═══════════════════════════════════╝
\end{Snippet}

If the "!" symbol is bound to the factorial function as defined on the
first picture, then pressing the ▶ or ▶▮ buttons will cause the
expression to be reduced by means of the substitution model of
procedure evaluation \cite{SICP}.

In this regard, the visual stepper in GRASP is similar to the stepper
available for the Beginner Student Language in the Dr Racket
programming environment \cite{Clements}.

What makes the stepper in GRASP different, is first that it uses the
representation of s-expressions as nested boxes, rather than text, and
second, that during the evaluation, subsequent steps of reductions are
smoothly morphed from one into another: the "!" will gradually morph
into the body of the factorial function, the "5" will triple itself,
and each of its occurrences will slide into the positions of the
occurrences of "n" in the body of the definition of factorial.

Eventually, we will get something that looks like this:

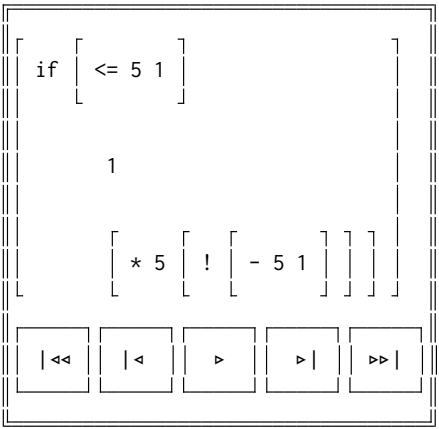
\begin{Snippet}
╔═══════════════════════════════════╗
║╭    ╭        ╮                 ╮  ║
║│ if │ <= 5 1 │                 │  ║
║│    ╰        ╯                 │  ║
║│                               │  ║
║│       1                       │  ║
║│                               │  ║
║│       ╭     ╭   ╭       ╮ ╮ ╮ │  ║
║│       │ * 5 │ ! │ - 5 1 │ │ │ │  ║
║╰       ╰     ╰   ╰       ╯ ╯ ╯ ╯  ║
║╭─────╮╭─────╮╭─────╮╭─────╮╭─────╮║
║│ ▮◀◀ ││ ▮◀  ││  ▶  ││  ▶▮ ││ ▶▶▮ │║
║╰─────╯╰─────╯╰─────╯╰─────╯╰─────╯║
╚═══════════════════════════════════╝
\end{Snippet}

Now, on the next step, the expression \texttt{(<= 5 1)} will morph into the
value \texttt{\#false}:

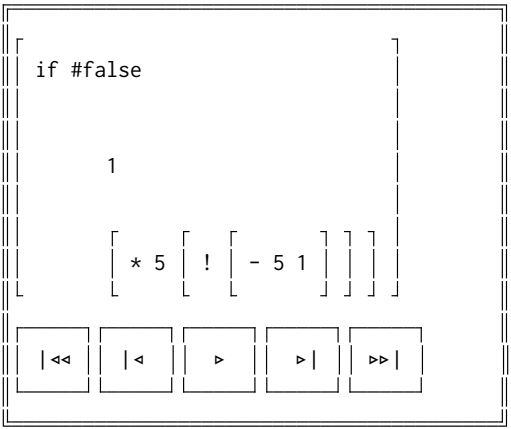
\begin{Snippet}
╔═════════════════════════════════════════╗
║╭                               ╮        ║
║│ if #false                     │        ║
║│                               │        ║
║│                               │        ║
║│       1                       │        ║
║│                               │        ║
║│       ╭     ╭   ╭       ╮ ╮ ╮ │        ║
║│       │ * 5 │ ! │ - 5 1 │ │ │ │        ║
║╰       ╰     ╰   ╰       ╯ ╯ ╯ ╯        ║
║╭─────╮╭─────╮╭─────╮╭─────╮╭─────╮      ║
║│ ▮◀◀ ││ ▮◀  ││  ▶  ││  ▶▮ ││ ▶▶▮ │      ║
║╰─────╯╰─────╯╰─────╯╰─────╯╰─────╯      ║
╚═════════════════════════════════════════╝
\end{Snippet}

Subsequently the whole expression will be replaced with the "else"
branch of the "if" expression: the content of the outermost expression
will fade away, and the expression \texttt{(* 5 (! (- 5 1)))} will
slide into the top left corner of the outer box.\footnote{A video
recording of this process (performed in both graphical and terminal
clients of GRASP) can be found at
\url{https://www.youtube.com/watch?v=wN8Fy5xTXeQ}, and the reader is
encouraged to watch it before proceeding through the next sections of
this paper.}

\section{The simplified model of reduction}

Computationally, visual stepper in GRASP is currently based on an
implementation of a small-step evaluator of an extended
lambda-calculus.

It uses a wrapper around Java's hash tables to represent
environments (herein called \texttt{Evalu\-ation\-Con\-text}s):

\begin{Snippet}
(define-object (EvaluationContext)
\end{Snippet}
\begin{Snippet}
  (define definitions ::java.util.Map
    (java.util.HashMap))
\end{Snippet}
\begin{Snippet}
  (define (value symbol)
    (cond ((definitions:contains-key symbol)
	   (definitions:get symbol))
	  (else
	   (error "undefined symbol: "symbol))))
\end{Snippet}
\begin{Snippet}
  (define (defines-macro? symbol)
    #f)
\end{Snippet}
\begin{Snippet}
  (define (defines? symbol)
    (definitions:contains-key symbol))
\end{Snippet}
\begin{Snippet}
  (define (define! name value)
    (definitions:put name value))
\end{Snippet}
\begin{Snippet}
  (define (primitive? symbol)
    (and (definitions:contains-key symbol)
	 (let ((value (definitions:get symbol)))
	   (procedure? value)))))
\end{Snippet}

In order to be able to use some of the Scheme's primitive
procedures, one has to export them to the \texttt{default-context}:

\begin{Snippet}
(define default-context ::EvaluationContext
  (EvaluationContext))
\end{Snippet}

If we wish to be able to use some of the primitive Scheme
functions, we need to make them available to the evaluator:

\begin{Snippet}
(default-context:define! '+ +)
(default-context:define! '- -)
(default-context:define! '* *)
(default-context:define! '/ /)
(default-context:define! '< <)
(default-context:define! '<= <=)
(default-context:define! '> >)
(default-context:define! '>= >=)
(default-context:define! '= =)
(default-context:define! 'eq? eq?)		
(default-context:define! 'eqv? eqv?)
\end{Snippet}

The main interface to the evaluator, i.e. the \texttt{reduce} procedure,
dispatches on the language's primitive forms: the \texttt{if} expression,
the \texttt{lambda} expression, the \texttt{quote} operator, combinations
and simple values (i.e. symbols and literals). The function
is designed so that it performs a single reduction in every step:

\begin{Snippet}
(define (reduce expression #!optional (context::EvaluationContext default-context))
  (match expression
    (`(if #f ,then ,else)
     else)
\end{Snippet}
\begin{Snippet}
    (`(if ,test ,then ,else)
     (let ((test* (reduce test context)))
       (if (equal? test test*)
	   then
	   `(if ,test* ,then ,else))))
\end{Snippet}
\begin{Snippet}
    (`(lambda ,args ,body)
     expression)
\end{Snippet}
\begin{Snippet}
    (`(quote ,_)
     expression)
\end{Snippet}
\begin{Snippet}
    (`(,operator . ,operands)
     (if (and (symbol? operator)
	      (context:defines-macro? operator))
	 (error "Macros not supported (yet)")
	 (let ((operands* (reduce-operands operands context)))
	   (if (isnt operands equal? operands*)
	       `(,operator . ,operands*)
	       (match operator
		 (,@symbol?
\end{Snippet}
\begin{Snippet}
		  (cond ((context:primitive? operator)
			 (apply (context:value operator) operands))
\end{Snippet}
\begin{Snippet}
			((context:defines? operator)
			 (reduce `(,(context:value operator) . ,operands)
				 context))
\end{Snippet}
\begin{Snippet}
			(else
			 `(,operator . ,operands))))
\end{Snippet}
\begin{Snippet}                                 
		 (`(lambda ,args ,body)
		  (substitute args #;with operands #;in body))
\end{Snippet}
\begin{Snippet}
		 (`(,_ . ,_)
		  (let ((operator* (reduce operator context)))
		    `(,operator* . ,operands)))
\end{Snippet}
\begin{Snippet}
		 (_
		  `(,operator . ,operands)))))))
\end{Snippet}
\begin{Snippet}
    (_
     (if (and (symbol? expression)
	      (context:defines? expression))
	 (context:value expression)
	 expression))))
\end{Snippet}

The \texttt{reduce-operands} procedure tries to reduce each of its
operands, from left to right:

\begin{Snippet}
(define (reduce-operands operands #!optional (context::EvaluationContext 
                                                         default-context))
  (match operands
    (`(,first . ,rest)
     (let ((first* (reduce first context)))
       (if (equal? first first*)
	   `(,first . ,(reduce-operands rest context))
	   `(,first* . ,rest))))
\end{Snippet}
\begin{Snippet}
    ('()
     '())
\end{Snippet}
\begin{Snippet}
    (_
     (reduce operands context))))
\end{Snippet}

The \texttt{substitute} procedure replaces all free occurrences of variables
with corresponding values in the given expression, accounting for
the \texttt{quote} operator (which precludes its argument from being evaluated)
and the \texttt{lambda} operator (which can shadow some of the free variables):

\begin{Snippet}
(define (substitute variables #;with values #;in expression)
  (match expression
\end{Snippet}
\begin{Snippet}
    (`(quote ,_)
     expression)
\end{Snippet}
\begin{Snippet}
    (`(lambda ,args ,body)
     (let-values (((variables* values*) (only. (isnt _ in. args) variables values)))
       `(lambda ,args
	  ,(substitute variables* #;with values* #;in body))))
\end{Snippet}
\begin{Snippet}
    (`(,operator . ,operands)
     `(,(substitute variables #;with values #;in operator)
       . ,(substitute variables #;with values #;in operands)))
\end{Snippet}
\begin{Snippet}
    (_
     (if (symbol? expression)
	 (counterpart #;of expression #;from variables #;in values)
	 expression))))
\end{Snippet}

Picking a counterpart of a variable given a list of variables and
values requires going through both lists simultaneously, until we run
into the variable that we're looking for (we also want to handle
dotted lists properly):

\begin{Snippet}
(define (counterpart #;of variable #;from variables #;in values)
  (match variables
    (`(,,variable . ,_)
     (let ((result (car values)))
       (if (self-evaluating? result)
	   result
	   `',result)))
\end{Snippet}
\begin{Snippet}
    (,variable
     `',values)
\end{Snippet}
\begin{Snippet}
    (`(,_ . ,rest)
     (counterpart #;of variable #;from rest #;in (cdr values)))
\end{Snippet}
\begin{Snippet}
    (_
     variable)))
\end{Snippet}

Values that are not self-evaluating are wrapped around in the
\texttt{quote} operator to prevent their further evaluation.

\section{Requirements for the visual stepper}

The stepper presented in the previous section used the classical
cons-cells, symbols and literals to represent expressions. All it did
was performing substitution in nested lists.

However, this is insufficient for the purpose of the visual evaluator
presented at the beginning of this paper. In addition to simply obtaining
new expressions, we also need to track the origins of the components
of its sub-expressions. Consider the reduction from the expression

\begin{Snippet}
(! 1)
\end{Snippet}

to

\begin{Snippet}
(if (<= 1 1)
    1
   (* 1 (! (- 1 1))))
\end{Snippet}

There are six occurrences of \texttt{1} in this expression, but only
three of them originate from argument substitution. Therefore, we need
to be able to track their identity using other means than the equality
predicates that are provided by Scheme.

Moreover, cons-cells themselves carry no information about line breaks
and indentation structure. This limitation has traditionally been
circumvented by pretty-printing, which can be confusing when the
indentation of the expression obtained from substitution changes
compared to the original expression. The source code can also contain
comments, and it can be desirable to preserve them in the process of
substitution.

\section{The representation of expressions in GRASP}

Some of the requirements from the previous section are already
satisfied by the representation of expressions that was developed for
representing documents in GRASP.\footnote{We do not claim, that the described
representation of expressions is particularly good, and we are open to
better alternatives.}

GRASP represents documents by subclassing the \texttt{pair} class
provided by Kawa. Originally the reason for it was that Kawa defines
an \texttt{equal?}-like \texttt{equals} method on cons-cells, which
do not allow to use cons-cells' pointer equality (\texttt{eq?}-like)
in the context of hash tables.

Initially GRASP used a number of hash tables, named
\texttt{pre\-/head\-/space}, \texttt{post\-/head\-/space},
\texttt{pre\-/tail\-/space} and \texttt{post\-/tail\-/space} for
representing spaces and comments between elements of the list.  It
also used additional hash tables called \texttt{null\-/head\-/space}
and \texttt{null\-/tail\-/space} to represent spaces contained
within empty lists.

There was a problem with editable representation of symbols:
given that the Scheme's \texttt{eq?} corresponds directly to the object
identity on the JVM, it was impossible to modify only a single occurrence
of a symbol, leaving the remaining ones intact. Moreover, it is
impossible to change object's type in run time, and in Scheme even some
very similar expressions (such as 1 and 1-) have different types (a
number and a symbol, respectively).

Therefore a new class called \texttt{Atom} was devised, that contained
an editable representation of atoms. A (SRFI-39-like) parameter called
\texttt{cell\-/access\-/mode} was introduced, and the \texttt{getCar}
and \texttt{getCdr} methods of the \texttt{cons} cell were overridden,
so that if the value of \texttt{(cell\-/access\-/mode)} was
\texttt{Cell\-Access\-Mode\-:Editing}, they would be returning
\texttt{Atom} objects, and otherwise if the parameter's value was
\texttt{Cell\-Access\-Mode\-:Evaluating}, they would return the parsed
content of \texttt{Atom} object's internal buffer.

Overriding the accessor methods also allowed to solve the problem with
the lack of identity of empty lists, and an object called
\texttt{Empty\-List\-Proxy} was introduced which held the internal
space of various instances of empty list. This allowed to remove the
\texttt{null\-/head\-/space} and \texttt{null\-/tail\-/space} hash
tables. Furthermore, some of the remaining tables were moved from hash
tables to the property list of the \texttt{cons} object with the hope
of optimizing the performance.

GRASP uses this representation to this day, although in retrospect
having two different access modes turned out to be very confusing, and
it would probably be better to have a different structure for editing
and a different one for evaluation, and conversion functions that
would transform between those two representations.

\section{The extended model of reduction}

The extended variant of the \texttt{reduce} function will take two additional
arguments. One of them, called \texttt{progeny}, will be a mutable hash-table
that maps a source element to all the sub-expressions that were created
by substituting that element with them. The second additional
argument, called \texttt{origin}, will map the other way around, from an
expression to all the expressions that were used to create it.

Although according to the reduction rules of lambda-calculus every
expression can have at most one origin element, we will represent both
tables as mappings from an element to a list of elements.

By default, the hash table of some element will return a list
containing only that element (which essentially means that by default
every element is its own origin/progeny).

Otherwise, an element can have many elements in its progeny list.
This corresponds to the argument substitution of arguments with
values.  It is also possible for an expression to have an empty
progeny list, which means that the expression disappears in the course
of reduction.

Both tables are populated with data as the reduction proceeds.  They
are also returned as additional values from the \texttt{reduce} function.

\begin{Snippet}
(define (reduce expression 
                #!optional 
                (origin::(!maps (Element) to: (list-of Element))
			 (property (e::Element)::(list-of Element)
				   (recons e '())))
		(progeny::(!maps (Element) to: (list-of Element))
			  (property (e::Element)::(list-of Element)
				    (recons e '())))
		#!key
		(context::EvaluationContext (default-context)))
\end{Snippet}
\begin{Snippet}
  (define (mark-origin! newborn parent)
    (set! (origin newborn) (recons parent '()))
    (set! (progeny parent) (recons newborn '())))
\end{Snippet}
\begin{Snippet}
  (define (add-origin! newborn parent)
    (and-let* ((`(,default) (origin newborn))
	       ((eq? newborn default)))
      (set! (origin newborn) '()))
    (and-let* ((`(,default) (progeny parent))
	       ((eq? parent default)))
      (set! (progeny parent) '()))
    (unless (any (is _ eq? parent) (origin newborn))
      (set! (origin newborn) (cons parent (origin newborn))))
    (unless (any (is _ eq? newborn) (progeny parent))
      (set! (progeny parent) (cons newborn (progeny parent)))))
\end{Snippet}
\begin{Snippet}
  (define (dissolve! item #!key (when? ::predicate
		                       (lambda (item)
		                         (and-let* ((`(,i) (progeny item))
			                            ((eq? i item)))))))
    (when (when? item)
      (for child in (progeny item)
	(set! (origin child) (only (isnt _ eq? item) (origin child))))
      (set! (progeny item) '()))
\end{Snippet}
\begin{Snippet}
    (when (gnu.lists.LList? item)
      (traverse item doing: (lambda (e::Element t::Traversal)
	                      (dissolve! e when?: when?)))))
\end{Snippet}
\begin{Snippet}
  (define (eradicate! item #!key (when? ::predicate
					(lambda (item)
					  (and-let* ((`(,i) (origin item))
						     ((eq? i item)))))))
    (when (when? item)
      (for child in (origin item)
	(set! (progeny child) (only (isnt _ eq? item) (progeny child))))     
      (set! (origin item) '()))
\end{Snippet}
\begin{Snippet}
    (when (gnu.lists.LList? item)
      (traverse item doing: (lambda (e::Element t::Traversal)
                              (eradicate! e when?: when?)))))
\end{Snippet}
\begin{Snippet}
  (define (substitute variables #;with values #;in expression)
\end{Snippet}
\begin{Snippet}
    (match expression
      (`(quote ,_)
       expression)
\end{Snippet}
\begin{Snippet}
      (`(lambda ,args ,body)
       (let*-values (((variables* values*) (only. (isnt _ in. args) variables values))
		     ((result) (cons* 
                                (car expression)
                                args
		                (substitute variables* #;with values* #;in body))))
	 (copy-properties cell-display-properties (cdr expression) (cdr result))
	 (copy-properties cell-display-properties expression result)
	 result))
\end{Snippet}
\begin{Snippet}
      (`(,operator . ,operands)
       (let ((result (cons (substitute variables #;with values #;in operator)
			   (substitute variables #;with values #;in operands))))
	 (mark-origin! result expression)
	 (copy-properties cell-display-properties expression result)))
\end{Snippet}
\begin{Snippet}
      (_
       (if (Atom? expression)
	   (counterpart #;of expression #;from variables #;in values)
	   expression))))
\end{Snippet}
\begin{Snippet}
  (define (counterpart #;of variable #;from variables #;in values)
    (match variables
\end{Snippet}
\begin{Snippet}
      (`(,,variable . ,_)
       (let* ((result (deep-copy (car values)))
	      (result (if (self-evaluating? result)
			  result
			  (cons (Atom "quote") result))))
	 (eradicate! result when?: always)
	 (add-origin! result (car variables))
	 result))
\end{Snippet}
\begin{Snippet}
      (,variable
       (let ((result (cons (Atom "quote") (copy values))))
	 (add-origin! result variable)
	 result))
\end{Snippet}
\begin{Snippet}
      (`(,_ . ,rest)
       (counterpart #;of variable #;from rest #;in (cdr values)))
\end{Snippet}
\begin{Snippet}
      (_
       variable)))
\end{Snippet}
\begin{Snippet}
  (define (reduce-operands operands)
    (match operands
\end{Snippet}
\begin{Snippet}
      (`(,first . ,rest)
       (let ((first* (reduce first)))
	 (if (match/equal? first first*)
	     (let ((result (cons first (reduce-operands rest))))
	       (mark-origin! result operands)
	       (copy-properties cell-display-properties operands result))
\end{Snippet}
\begin{Snippet}
	     (let ((result (cons first* rest)))
	       (mark-origin! result operands)
	       (copy-properties cell-display-properties operands result)))))
\end{Snippet}
\begin{Snippet}
      (`()
       operands)
\end{Snippet}
\begin{Snippet}
      (_
       (reduce operands))))
\end{Snippet}
\begin{Snippet}
  (define (deep-copy expression)
    (match expression
\end{Snippet}
\begin{Snippet}
      (`(,h . ,t)
       (let ((result (cons (deep-copy h) (deep-copy t))))
	 (mark-origin! result expression)
	 (copy-properties cell-display-properties expression result)
	 result))
\end{Snippet}
\begin{Snippet}
      (_
       (let ((result (copy expression)))
	 (mark-origin! result expression)
	 result))))
\end{Snippet}
\begin{Snippet}
  (define (transfer-heritage! args vals)
    (match args
\end{Snippet}
\begin{Snippet}
      (`(,arg . ,args*)
       (let ((val (car vals))
	     (vals* (cdr vals))
	     (children (progeny arg)))
	 (set! (progeny val) children)
	 (for p in children
	   (set! (car (origin p)) val))
	 (transfer-heritage! args* vals*)))
\end{Snippet}
\begin{Snippet}
      ('()
       (values))
\end{Snippet}
\begin{Snippet}
      (arg
       (let ((children (progeny arg)))
	 (set! (progeny vals) children)
	 (for p in children
	   (set! (car (origin p)) vals))))))
\end{Snippet}
\begin{Snippet}
  (define (reduce expression)
    (match expression
\end{Snippet}
\begin{Snippet}
      (`(if #f ,then ,else)
       (dissolve! expression)
       (let ((result (deep-copy else)))
	 (mark-origin! result else)
	 result))
\end{Snippet}
\begin{Snippet}
      (`(if ,test ,then ,else)
       (let ((test* (reduce test))
	     (if* (car expression)))
	 (cond ((match/equal? test test*)
		(dissolve! expression)
		(let ((result (deep-copy then)))
		  (mark-origin! result then)
		  result))
\end{Snippet}
\begin{Snippet}
	       (else
		(let ((result (cons* if* test* then else '())))
		  (mark-origin! result expression)
		  (mark-origin! test* test)
		  (copy-properties* cell-display-properties expression result)
		  result)))))
\end{Snippet}
\begin{Snippet}
      (`(lambda ,args ,body)
       expression)
\end{Snippet}
\begin{Snippet}
      (`(quote ,_)
       expression)
\end{Snippet}
\begin{Snippet}
      (`(,operator . ,operands)
       (if (and (Atom? operator)
		(context:defines-macro? operator))
	   (error "Macros not supported (yet)")
\end{Snippet}
\begin{Snippet}
	   (let ((operands* (reduce-operands operands)))
	     (if (isnt operands match/equal? operands*)
		 (let* ((operator* (copy operator))
			(result (cons operator* operands*)))
		   (mark-origin! operator* operator)
		   (mark-origin! operands* operands)
		   (mark-origin! result expression)
		   (copy-properties cell-display-properties expression result))
\end{Snippet}
\begin{Snippet}
		 (match operator
		   (,@Atom?		    
		    (cond ((context:primitive? operator)
			   (let ((result 
                                  (grasp
			           (parameterize ((cell-access-mode 
                                                   CellAccessMode:Evaluating))
				     (apply (context:value operator)
					    (map (lambda (x) x) operands))))))
			     (mark-origin! result expression)
			     result))
\end{Snippet}
\begin{Snippet}
			  ((context:defines? operator)
			   (let ((operator* (context:value operator)))
			     (match operator*
\end{Snippet}
\begin{Snippet}
			       (`(lambda ,args ,body)
				(let ((result (substitute args #;with operands
                                                          #;in body)))
				  (transfer-heritage! args operands)
				  (dissolve! expression)
				  (mark-origin! result operator)
				  result))
\end{Snippet}
\begin{Snippet}
			       (_
				`(,operator* . ,operands)))))
\end{Snippet}
\begin{Snippet}
			  (else
			   expression)))
\end{Snippet}
\begin{Snippet}
		   (`(lambda ,args ,body)
		    (dissolve! expression)
		    (let ((result (substitute args #;with operands #;in body)))
		      result))
\end{Snippet}
\begin{Snippet}
		   (`(,_ . ,_)
		    (let* ((operator* (reduce operator))
			   (result (cons operator* operands)))
		      (mark-origin! result expression)
		      (mark-origin! operator* operator)
		      (copy-properties cell-display-properties expression result)))
\end{Snippet}
\begin{Snippet}
		   (_
		    expression))))))
\end{Snippet}
\begin{Snippet}
      (_
       (if (and (Atom? expression)
		(context:defines? expression))
	   (let ((result (copy (context:value expression))))
	     (dissolve! expression)
	     (mark-origin! result expression)
	     result)
\end{Snippet}
\begin{Snippet}
	   expression))))
\end{Snippet}
\begin{Snippet}
  (values (reduce expression)
	  origin
	  progeny))
\end{Snippet}

The \texttt{traverse} function is used for iterating over subsequent
elements in the document, where even elements are spaces/comments, and
odd elements are actual data.  The
\texttt{cell\-/display\-/pro\-per\-ties} variable points to a list with
references to \texttt{pre\-/head\-/space},
\texttt{post\-/head\-/space}, \texttt{pre\-/tail\-/space} and
\texttt{post\-/tail\-/space}.

\section{The rendering subsystem in GRASP}

As we mentioned at the beginning, GRASP is available for desktop, terminal
and Android. Since rendering and input handling APIs are incompatible
between those systems, GRASP provides an interface called \texttt{Painter},
which is responsible for drawing things on the screen and measuring their
sizes.

The \texttt{Painter} interface is not simple, mainly because it needs
to compensate for the peculiarities of rendering graphics into the
terminal.

So in addition to drawing things such as lines and rectangles, it
needs to know how to write things like buttons and parentheses (box
edges).

While the extension system of GRASP should allow for drawing a wide
variety of things, at the current stage of the project it can turn out
that the functions required for rendering those things are not
provided by the \texttt{Painter} interface, and that the author of
extension needs to add a new set of functions to that interface (which
also means providing three implementations). Although this solution is
not ideal, the hope is that over time the interface will accumulate
enough functions for drawing all the desired extensions.

In particular, for the purpose of the visual evaluator, we have
extended the \texttt{Painter} interface with the following methods:

\begin{Snippet}
(with-intensity i::float action::(maps () to: void))::void
(with-stretch horizontal::float vertical::float action::(maps () to: void))::void
\end{Snippet}

The first of those methods, \texttt{with-intensity}, decreases the
intensity of rendered objects by multiplying it by \texttt{i}, which
is meant to be a number between 0 and 1 (where 1 means full intensity,
and 0 means that an object is invisible)

The second one, called \texttt{with-stretch}, takes two floating point
numbers and scales the rendered object along the horizontal and the
vertical axis, accordingly.

The visual stepper will also need another method that is used
extensively for rendering GRASP documents, namely \texttt{translate!},
which shifts the origin of the drawing, and is used by the following
macro:

\begin{Snippet}
(define-syntax-rule (with-translation (x y)
		      . actions)
  (let ((x! ::real x)
        (y! ::real y))
    (painter:translate! x! y!)
    (try-finally
     (begin . actions)
     (painter:translate! (- x!) (- y!)))))
\end{Snippet}

where \texttt{painter} is a global variable that holds a reference
to the application's implementation of the \texttt{Painter} interface.

\section{Rendering transitions between expressions}

The most spectacular part of the visual stepper are transitions. They
are expressed using the \texttt{Morph} object, which -- among other
things -- contains the \texttt{progress} property, which is a real
number between 0 and 1, where 0 means that we should only render the
source expression, while 1 means that we should only render the target
expression.

For every other value in that range, we should get an interpolation
between those two expressions.

The \texttt{Morph} class is defined in the following way:

\begin{Snippet}
(define-object (Morph initial::Tile final::Tile
		      origin::(maps (Element) to: (list-of Element))
		      progeny::(maps (Element) to: (list-of Element)))
  ::Enchanted
\end{Snippet}
\begin{Snippet}
  (define progress ::float 0.0)
\end{Snippet}
\begin{Snippet}
  (define initial-position ::(maps (Element) to: Position)
    (measure-positions! initial))
\end{Snippet}
\begin{Snippet}
  (define initial-extent ::Extent
    (extent+ initial))
\end{Snippet}
\begin{Snippet}
  (define final-position ::(maps (Element) to: Position)
    (measure-positions! final))
\end{Snippet}
\begin{Snippet}
  (define final-extent ::Extent
    (extent+ final))
\end{Snippet}
\begin{Snippet}
  (define maximum-extent ::Extent
    (Extent width: (max initial-extent:width final-extent:width)
	    height: (max initial-extent:height final-extent:height)))
\end{Snippet}
\begin{Snippet}
  (define (extent) ::Extent maximum-extent)
\end{Snippet}
\begin{Snippet}
  (define shift ::(maps (Element) to: Position)
    (property+ (element::Element)::Position
	       (Position left: 0 top: 0)))
\end{Snippet}
\begin{Snippet}
  (define (draw! context::Cursor)::void
    (cond ((is progress <= 0.5) 
 	   (draw-tween! final origin final-position initial-position progress)
           (draw-tween! initial progeny initial-position final-position 
                        (- 1.0 progress)))
	  (else
           (draw-tween! initial progeny initial-position final-position 
                        (- 1.0 progress))
           (draw-tween! final origin final-position initial-position progress))))
  (Magic))
\end{Snippet}

As we can see, it takes two arguments -- the \texttt{initial}
expression, the \texttt{final} expression and the two maps returned by
the \texttt{reduce} function.

Upon initialization, it measures the positions of all sub-expressions
of the \texttt{initial} and \texttt{final} expressions.

The class is defined as a subclass of \texttt{Magic} that defines the
\texttt{Enchanted} interface, which is required by the extension
system of GRASP.

The \texttt{draw!} method is defined so that if \texttt{progress} is
no greater than 0.5, then we render the final expression as the
background, and then -- on top of it -- we draw the initial expression
in the foreground.  But once the progress of 0.5 is exceeded, we first
draw the initial expression, and then we draw the final expression on
top of it.

This allows to achieve satisfying visual effects even in the terminal
client of GRASP, which does not provide any mechanisms for
transparency.

The \texttt{draw-tween!} function has to support two cases: the first
one is when the list of the rendered expression's counterparts is
empty. In such a case, we want this expression to dissolve into
background.

Otherwise we want to morph the expression into each of its counterparts.

\begin{Snippet}
(define (draw-tween! expression::Element
		     counterparts::(maps (Element) to: (list-of Element))
		     source-position::(maps (Element) to: Position)
		     target-position::(maps (Element) to: Position)
		     intensity::float
		     #!key (only-with-relatives ::boolean #f))
  ::void
\end{Snippet}
\begin{Snippet}
  (let ((links (counterparts expression)))
    (cond
     ((empty? links)
      (draw-emerging! expression (source-position expression) intensity)
      (when (gnu.lists.LList? expression)
        (traverse
         expression
         doing:
         (lambda (sub::Element t::Traversal)
	   (draw-tween! sub counterparts source-position target-position
	                intensity only-with-relatives: only-with-relatives)))))
\end{Snippet}
\begin{Snippet}
    (else
     (for x in links
       (draw-morph! expression x counterparts source-position target-position
		    intensity only-with-relatives: only-with-relatives))))))
\end{Snippet}

When it comes to morphing, we need to do three things. First, we need
to find the interpolation between the positions of the source and the
target expression. Second, we need to stretch the rendered expressions
to make their sizes an interpolation between the source and the target
expressions.  Third, we need to adjust the intensity of the source and
the target expressions to make the effect of fading from one
expression to another.

Thus, the \texttt{draw-morph!} procedure is defined in the following way:

\begin{Snippet}
(define (draw-morph! foreground::Element background::Element
		     counterparts::(maps (Element) to: (list-of Element))
		     source-position::(maps (Element) to: Position)
		     target-position::(maps (Element) to: Position)
		     progress::float
		     #!key (only-with-relatives ::boolean #f))
  ::void
\end{Snippet}
\begin{Snippet}
  (let* ((p0 ::Position (source-position foreground))
	 (p1 ::Position (target-position background))
	 (left ::real (linear-interpolation from: p0:left to: p1:left
		                            at: (- 1 progress)))
	 (top ::real (linear-interpolation from: p0:top to: p1:top
		                           at: (- 1 progress))))
\end{Snippet}
\begin{Snippet}
    (cond
     ((match/equal? foreground background)
      ;; here we just draw the foreground
      ;; with full intensity
      (unless (and only-with-relatives (eq? foreground background))
	(with-translation (left top)
	  (draw! foreground))))
\end{Snippet}
\begin{Snippet}
     ((or (isnt foreground Tile?)
	  (isnt background Tile?))
      ;; at least one of the elements is (presumably)
      ;; a space, so the only way we can morph them
      ;; is by fading
      (with-translation (left top)
	(painter:with-intensity (- 1.0 progress)
	  (lambda ()
	    (draw! background)))
	(painter:with-intensity progress
	  (lambda ()
	    (draw! foreground)))))
\end{Snippet}
\begin{Snippet}
     ((and (gnu.lists.LList? foreground)
	   (gnu.lists.LList? background))
      (let* ((e0 ::Extent (extent+ foreground))
	     (e1 ::Extent (extent+ background))
	     (width ::real (linear-interpolation from: e0:width to: e1:width
						 at: (- 1 progress)))
	     (height ::real (linear-interpolation from: e0:height to: e1:height
						  at: (- 1 progress))))
\end{Snippet}
\begin{Snippet}
	(unless only-with-relatives
	  (with-translation (left top)
	    (painter:draw-box! width height '())))
\end{Snippet}
\begin{Snippet}
	(traverse
	 foreground
	 doing:
	 (lambda (item::Element t::Traversal)
	   (draw-tween! item counterparts source-position target-position
			progress only-with-relatives: only-with-relatives)))))
\end{Snippet}
\begin{Snippet}
     ((and (Tile? foreground)
	   (Tile? background))
      (let* ((e0 ::Extent (extent+ foreground))
	     (e1 ::Extent (extent+ background))
	     (width ::real (linear-interpolation from: e0:width to: e1:width
						 at: (- 1 progress)))
	     (height ::real (linear-interpolation from: e0:height to: e1:height
	 					  at: (- 1 progress))))
\end{Snippet}
\begin{Snippet}
	(with-translation (left top)
	  (painter:with-intensity (- 1.0 progress)
	    (lambda ()
	      (painter:with-stretch (/ width e1:width)
                  (/ height e1:height)
		(lambda ()
		  (draw! background)))))
\end{Snippet}
\begin{Snippet}
	  (painter:with-intensity progress
	    (lambda ()
	      (painter:with-stretch (/ width e0:width)
                  (/ height e0:height)
		(lambda ()
		  (draw! foreground)))))))
\end{Snippet}
\begin{Snippet}
      (when (gnu.lists.LList? foreground)
	(traverse foreground
		  doing:
		  (lambda (element::Element traverse::Traversal)
		    (draw-tween! element counterparts
				 source-position
				 target-position
				 progress
				 only-with-relatives: #t)))))
     )))
\end{Snippet}

The \texttt{draw-emerging!} supplementary procedure is defined as

\begin{Snippet}
(define (draw-emerging! expression::Element p::Position
			intensity::float)
  ::void
  (painter:with-intensity intensity
    (lambda ()
      (with-translation (p:left p:top)
	(if (gnu.lists.LList? expression)
	    (let ((outer ::Extent (extent+ expression)))
	      (painter:draw-box! outer:width outer:height '()))
	    (draw! expression))))))
\end{Snippet}

\section{The extension mechanism of GRASP}

The \texttt{Stepper} extension is defined using the following code:

\begin{Snippet}
(define-simple-extension (Stepper expression::Tile)
  (PlayerWithControls (ExpressionReducer expression)))
\end{Snippet}

and made visible to the extension system in the following way (where
\texttt{object} is a special form provided by Kawa Scheme for creating
anonymous classes):

\begin{Snippet}
(set! (extension 'Stepper)
      (object (Extension)
	((enchant source::cons)::Enchanted
	 (parameterize ((cell-access-mode CellAccessMode:Editing))
	   (or (and-let* ((`(Stepper ,expression) source))
		 (Stepper expression)))
	     (WARN "Unable to create Stepper from "source)))))
\end{Snippet}

It therefore plays a role similar to the \texttt{quote} operator, in that
it passes \texttt{expression} directly to the \texttt{Stepper} constructor
without evaluating it.

The \texttt{define-simple-extension} form is just a wrapper that compensates
for some shortcomings of the object system implemented for GRASP:

\begin{Snippet}
(define-syntax define-simple-extension
  (syntax-rules ()
    ((_ (name args ...) body)
     (define-object (name args ...)::Enchanted 
       (define (typename)::String
	 (symbol->string 'name))
\end{Snippet}
\begin{Snippet}
       (define (value)::cons
	 (cons (Atom (symbol->string 'name))
	       (dropping-type-signatures list*
					 disenchanted
					 (args ...)
					 ())))
       (SimpleExtension body)))))
\end{Snippet}

where the \texttt{typename} method is simply used for retrieving the
name of a type, and the \texttt{value} method is used for transforming
the enchanted object back to its source form. The
\texttt{SimpleExtension} superclass defines an object that passes all
the touch events to its argument (which must be an enchanted object).

The \texttt{PlayerWithControls} procedure is defines in the following way:

\begin{Snippet}
(define (PlayerWithControls player::Player)::Enchanted
  (bordered
   (below
    player
    (beside
      (Button label: "▮◀◀" action: (lambda () (player:rewind!)))
      (Button label: "▮◀ " action: (lambda () (player:back!)))
      (Button label: " ▶ " action: (lambda () (player:play!)))
      (Button label: " ▶▮" action: (lambda () (player:next!)))
      (Button label: "▶▶▮" action: (lambda () (player:fast-forward!))))
    )))
\end{Snippet}

where \texttt{bordered}, \texttt{below} and \texttt{beside} procedures
are extension combinators.

The \texttt{Player} interface is defined as

\begin{Snippet}
(define-interface Player (Enchanted Playable Animation))
\end{Snippet}

where the \texttt{Playable} interface is defined as

\begin{verbatim}
(define-interface Playable ()
  (rewind!)::void
  (back!)::void
  (play!)::void
  (pause!)::void
  (next!)::void
  (fast-forward!)::void
  (playing?)::boolean)
\end{verbatim}

and \texttt{Animation} is something that can simply be ``advanced'' or
``pushed forward'':

\begin{Snippet}
(define-interface Animation ()
  (advance! timestep/ms::int)::boolean)
\end{Snippet}

The \texttt{Enchanted} interface is a composition of two more
fundamental interfaces defined by GRASP:

\begin{Snippet}
(define-interface Enchanted (Interactive ShadowedTile))
\end{Snippet}

The \texttt{Interactive} interface contains a number of methods that
can be used to interact with a document element:

\begin{verbatim}
(define-interface Interactive ()
  (tap! finger::byte #;at x::real y::real)::boolean
  (press! finger::byte #;at x::real y::real)::boolean
  (second-press! finger::byte #;at x::real y::real)::boolean
  (double-tap! finger::byte x::real y::real)::boolean
  (long-press! finger::byte x::real y::real)::boolean
  (key-typed! key-code::long context::Cursor)::boolean

  (scroll-up! left::real top::real)::boolean
  (scroll-down! left::real top::real)::boolean
  (scroll-left! left::real top::real)::boolean
  (scroll-right! left::real top::real)::boolean
  (zoom-in! left::real top::real)::boolean
  (zoom-out! left::real top::real)::boolean
  (rotate-left! left::real top::real)::boolean
  (rotate-right! left::real top::real)::boolean)
\end{verbatim}

The \texttt{ShadowedTile} interface, again, is a composition
of two simpler interfaces:

\begin{Snippet}
(define-interface ShadowedTile (Shadowed Tile))
\end{Snippet}

where \texttt{Shadowed} simply lets the object to be viewed
differently in the editing context and the evaluation context
(depending on the value of the \texttt{cell\-/access\-/mode}
parameter), and \texttt{Tile} is defined as

\begin{Snippet}
(define-interface Tile (Extensive Element))
\end{Snippet}

\texttt{Extensive} means that a thing has its \texttt{extent},
or simply \texttt{width} and \texttt{height}:

\begin{Snippet}
(define-type (Extent width: real := 0
                     height: real := 0))
  
(define-interface Extensive ()
  (extent)::Extent)
\end{Snippet}

\texttt{Element} is probably the most fundamental class in GRASP, and
it represents every visible component of the document, including
lists, atoms and spaces (with comments). In fact, spaces are the only
\texttt{Element}s in GRASP that aren't \texttt{Tile}s.

Covering the entire \texttt{Element} interface is beyond the scope of
this work. From our perspective, the most important aspect is that it
requires the \texttt{draw!} method that is used for executing the
rendering code.

\section{Limitations and future work}

Currently the visual stepper works only on a very limited subset of
the Scheme programming language, namely -- on purely functional
programs that operate on numbers. The ▶▶▮ button is a complete fake
and doesn't even pretend to do anything (other than looking nice and
symmetrical).

Some work needs to be done in order to support programs that operate
on lists, and on programs whose execution is non-deterministic.

It is known, that the substitution model of procedure evaluation is
not suitable to all classes of Scheme programs. It would therefore be
desirable to have a stepper that is able to choose an adequate
visualization method for particular programs.

Another thing is that the stepper is not meta-circular. It is
implemented in Kawa Scheme, which uses a Java-like object system built
around the notion of interfaces, and it is not obvious how such
programs should be visualized.

From the functional point of view, it would probably be desirable to be
able to point to the stepper, which reductions should be considered
primitive, and which ones should be tracked. There is also an issue
of macro expansion.

We believe that such a tool would be invaluable for learning new
code bases.

Another issue is the GRASP editor itself. While it admittedly does
look slick and impressive, the only impression that one can get from
actually trying to use it beyond toy examples, is irritation.

While GRASP was first demoed during the Scheme workshop in
2021 \cite{Godek2021}, and its author has certainly made a lot of
progress (including rewriting the editor from Java to Scheme, and
making it work on platforms other than Android), it still gives an
impression of a work-in-progress, rather than a well-polished project.

\section{(Un)related work}

The name ``GRASP'' doesn't seem to be particularly original in the
realm of software development tools. In the context of Scheme, the
name ``GRASP'' has been used by Franklyn Turbak in his 1986 master
thesis titled \textit{GRASP: A Visible and Manipulable Model for
  Procedural Programs} \cite{Turbak}. In the context of Java, there
exists a project called \textit{jGRASP}, which is \textit{a
  lightweight development environment, created specifically to provide
  automatic generation of software visualizations to improve the
  comprehensibility of software} \cite{jGRASP}.

While the descriptions of both these works may sound like they are
having similar goals to the project described in this paper, they are
not related in any direct way.

In some ways, GRASP is reminiscent of Andrea diSessa's Boxer
project \cite{Boxer}, which describes a new computational medium, as
well as Bochser \cite{Bochser}, which is a Boxer spin-off built around
the Scheme programming language.

The extension capabilities of GRASP are similar to the ones that can
be found in Hazel\footnote{\url{https://hazel.org}},
Polytope\footnote{\url{https://elliot.website/editor/}} and
Interactive Visual Syntax (as implemented for the Dr Racket
programming environment and a ClojureScript online IDE
\url{https://visr.pl}) \cite{Andersen}.

The work that is closest to GRASP's visual stepper is probably the
algebraic stepper that can be found in the dr Racket programming
environment \cite{Clements}. However, the latter uses a very different
implementation technique, based on the notion of \textit{continuation
  marks}, while the stepper in GRASP is currently based on a
substitution-based evaluator, reminescent of the ones that can be
found in \cite{Maritime}, \cite{Meyer} or \cite{Pedersen}.

\section*{Acknowledgements}

The author would like to express his gratitude to Pawel Pawlowski for
reviewing the first draft of this paper, and to countless souls on the
Internet for their spiritual support, including Scislaw Dercz,
Amirouche Bobekki, Andrew Blinn, Kartik Agaram, Oleksandr Kryvonos,
Jonas Winje, Jack Rusher, Tony Garnock-Jones, and the whole
\textit{Future of Coding} community.

\appendix
\section{The Kawa Scheme language and its use in GRASP}

As mentioned in the paper, GRASP is implemented in the Kawa Scheme programming language.
Kawa is a superset of Scheme, and it extends the core language with:
\begin{enumerate}
\item optional (checked) type annotations to variables and procedure return types
\item the ability to define new Java-style classes
\item interoperability with the features available on the JVM platform
\end{enumerate}

\subsection*{Type annotations}

The type annotations are marked with the \texttt{::} symbol. Kawa's
reader is designed in such a way that it will read two subsequent
colons as a separate token. Therefore the expression
\texttt{(call\-/with\-/input\-/string "(a::b)" read)} will evaluate
to a list of three elements: the symbol \texttt{a}, the symbol
\texttt{::} and the symbol \texttt{b}.

An example of an annotated function might look like this:

\begin{Snippet}
  (define (string-length s::string)::integer
     #| body of the definition omitted |#)
\end{Snippet}

The expressions in the type position are macro-expanded.

Kawa also provides a DSSSL-style (or Common Lisp-style) syntax for
defining optional arguments and keyword arguments, which additionally
allow to annotate types of such arguments.

\subsection*{Class definitions}

Another extension to the reader provided by Kawa is the colon
syntax. The evaluation of the expression
\texttt{(call\-/with\-/input\-/string "a:b" read)} produces the list
\texttt{(\$lookup\$ a (quasiquote b))}, and \texttt{\$lookup\$} is
defined as a method invocation or a property retrieval.

Kawa provides an interface for defining Java classes in the form of
the \texttt{define\-/simple\-/class} syntax, which is used in the
following way:

\begin{Snippet}
  (define-simple-class <class-name> (<super-classes> ...)
    <methods-and-properties>
    ...)
\end{Snippet}

where a slot definition in its simplest incarnation can have a form

\begin{Snippet}
  (<slot-name> ::<type> init-value: <value>)
\end{Snippet}

and a method definition has a form

\begin{Snippet}
  ((<method-name> <arguments>...)::<return-type> . <body>)
\end{Snippet}

If a method body consists of a single token, \texttt{\#!abstract},
then it is an abstract method, and a whole class becomes abstract.
A special method name, \texttt{*init*}, is reserved for defining
constructors.

\subsection*{GRASP wrappers for defining classes}

Kawa also provides its own object system built on top of
that of Java, which -- according to the manual -- allows for true
multiple inheritance, and is available through the
\texttt{define\-/class} form. However, this system is never used in
GRASP, which instead defines its own macros that wrap the
\texttt{define\-/simple\-/class} form around.

\subsubsection*{\texttt{\textbf{define-type}}} \ \\

The first such macro is \texttt{define\-/type} defined in the
\texttt{(language define\-/type)} module. It allows to define records
of form

\begin{Snippet}
  (define-type (TypeName field1: type1 
                         field2: type2 := initial-value))
\end{Snippet}

The record fields can be accessed using the colon notation, as in

\begin{Snippet}
  (define instance ::TypeName
    (TypeName field1: value1
              field2: value2))

  instance:field1 ; should return value1
\end{Snippet}

The \texttt{(language match)} module also contains the definition
of a \texttt{match} form that allows to pattern-match against records:

\begin{Snippet}
  (match instance
    ((TypeName field1: ,particular-value
               field2: any-value)
     ;; if instance has type TypeName, and its field1's value is
     ;; match/equal? to particular-value, the any-value identifier will
     ;; be bound to instance's field2 value in here
     ...)
    ...)          
\end{Snippet}

The records can additionally extend existing classes and implement
interface methods.

\subsubsection*{\texttt{\textbf{define-interface}}} \ \\

The \texttt{(language define\-/interface)} module defines a macro called
\texttt{define\-/interface}, which is used as

\begin{Snippet}
  (define-interface InterfaceName (SuperInterfaces ...)
    (method1-name args ...)::return-type1
    (method2-name other-args ...)::return-type2
    ...)
\end{Snippet}

Note that the last ellipsis is meant to generalize over triples, so
its semantics does not conform to that of the \texttt{syntax\-/rules}
pattern language.

\subsubsection*{\texttt{\textbf{define-object}}} \ \\

The third wrapper around \texttt{define\-/simple\-/class} is called
\texttt{define\-/object}, and is used as

\begin{Snippet}
  (define-object (ObjectName constructor-args...)::ImplementedInterface
  
    (define (method args ...)::return-type (body ...))
  
    (define property ::type init-expression)
  
    (SuperClass optional-constructor-args ...)

    initialization-code ...)
\end{Snippet}

This syntax imposes the following limitations on the way in which
the \texttt{define\-/object} system can be used:
\begin{itemize}
\item it only allows for a single constructor
\item it only allows for a single super-class constructor call
\item it only allows a class to implement a single interface
\end{itemize}

If we need to have a class that implements more than one interface,
an interface aggregate needs to be created first.

\end{document}